\documentclass[twocolumn,aps,floatfix,superscriptaddress]{revtex4-2}
\usepackage{amssymb,amsmath,bm,graphicx}
\usepackage{enumitem}

\begin{document}
\title{Nonreciprocal interactions give rise to fast cilium synchronisation in finite systems}
	\author{David J. Hickey}
	\affiliation{Max Planck Institute for Dynamics and Self-Organization (MPIDS), 37077 G\"ottingen, Germany}
	\author{Ramin Golestanian}
        \email{ramin.golestanian@ds.mpg.de}
	\affiliation{Max Planck Institute for Dynamics and Self-Organization (MPIDS), 37077 G\"ottingen, Germany}
	\affiliation{Rudolf Peierls Centre for Theoretical Physics, University of Oxford, Oxford OX1 3PU, United Kingdom}
	\author{Andrej Vilfan}
	\email{andrej.vilfan@ds.mpg.de}
	\affiliation{Max Planck Institute for Dynamics and Self-Organization (MPIDS), 37077 G\"ottingen, Germany}
	\affiliation{Jo\v{z}ef Stefan Institute, 1000 Ljubljana, Slovenia}
	\date{\today}

	\begin{abstract}
Motile cilia beat in an asymmetric fashion in order to propel the surrounding fluid. When many cilia are located on a surface, their beating can synchronise such that their phases form metachronal waves. Here, we computationally study a model where each cilium is represented as a spherical particle, moving along a tilted trajectory with a position-dependent active driving force and a position-dependent internal drag coefficient. The model thus takes into account all the essential broken symmetries of the ciliary beat. We show that taking into account the near-field hydrodynamic interactions, the effective coupling between cilia can become nonreciprocal: the phase of a cilium is more strongly affected by an adjacent cilium on one side than by a cilium at the same distance in the opposite direction. As a result, synchronisation starts from a seed at the edge of a group of cilia and propagates rapidly across the system, leading to a synchronisation time that scales proportionally to the linear dimension of the system. We show that a ciliary carpet is characterised by three different velocities: the velocity of fluid transport, the phase velocity of metachronal waves and the group velocity of order propagation. Unlike in systems with reciprocal coupling, boundary effects are not detrimental for synchronisation, but rather enable the formation of the initial seed. 
\end{abstract}

\maketitle

Motile cilia are hairlike organelles which can move under their own power in order to fulfil roles such as fluid pumping or mixing~\cite{brennen_fluid_1977}. They are nigh-ubiquitous in biological systems, being found on most eukaryotic cells~\cite{nachury_establishing_2019} including in the nervous system~\cite{faubel_cilia-based_2016}, the respiratory system~\cite{yaghi_airway_2016}, and the olfactory system~\cite{bhandawat_signaling_2010}. This makes them central to many open questions in biology, such as the precise mechanism behind the emergence of left-right differentiation during embryonic development~\cite{dasgupta_cilia_2016}. While the fascinating fluid dynamical questions involved in the dynamics of cilia and their biological function have been already highlighted by the pioneers of twentieth century fluid dynamics such as Ludwig Prandtl \cite{prandtl:1926} and G. I. Taylor \cite{taylor1951}, the subject of the collective properties of hydrodynamically active organelles at low Reynolds number continues to be an active field of research, particularly as a key component of the field of active matter  \cite{golestanian_hydrodynamic_2011}.

When many motile cilia are located on a surface at sufficient density, their beating can synchronise with a phase lag between neighbouring cilia. The resulting phase waves are called metachronal waves. It has been shown that metachronal co-ordination can lead to a high energetic efficiency of swimming or fluid transport \cite{osterman_finding_2011, elgeti_emergence_2013}, and that metachronal waves may reduce collisions between cilia, further raising pumping speed \cite{ringers_novel_2023}. Moreover, the coordination has been shown to be beneficiary for the efficiency of the chemosensory function of motile cilia  \cite{hickey_ciliary_2021}. Metachronal waves are found in many different organisms and systems. For example, \textit{Paramecium} uses metachronally coordinated cilia to swim~\cite{funfak_paramecium_2015}, as well as to feed~\cite{funfak_paramecium_2015}. Indeed, \textit{Paramecium}'s swimming efficiency is close to the maximum possible efficiency for an organism with cilia of that length~\cite{osterman_finding_2011}. Metachronal waves are found in other systems, such as the multicellular colony \textit{Volvox} \cite{brumley_hydrodynamic_2012} or cilia in the respiratory tract~\cite{yaghi_airway_2016} where their pumping efficiency is important for moving mucus \cite{chateau_why_2019}. Metachronal coordination also appears in animals (e.g., krill) at larger length scales with very different coordination mechanisms \cite{byron_metachronal_2021}.

Metachronal waves can be classified according to the direction of the wave propagation, depending on how the phase velocity of the wave compares to the direction of fluid transport. When these two directions are parallel, the metachronal wave is said to be symplectic. If they are antiparallel, the wave is called antiplectic~\cite{knight-jones_relations_1954}. Other wave directions are classified as dexioplectic or leoplectic. 

The fact that a pair of hydrodynamically interacting cilia or flagella can synchronise their cycles, even when belonging to two separate organisms \cite{goldstein_noise_2009}, suggests that hydrodynamic coupling alone can be sufficient to explain the emergence of metachronal waves. Nevertheless, some studies also point to the additional role of intracellular linkages \cite{wan_coordinated_2016, quaranta_hydrodynamics_2015, liu_transitions_2018}. 
In fact, the metachronal waves in \textit{Paramecium} can preserve synchrony across the wall of a micropipette that isolates them hydrodynamically \cite{narematsu_ciliary_2015}. 

A fundamental problem in understanding synchronisation via hydrodynamic interactions is the reversible nature of the Stokesian hydrodynamics, i.e.\ the fact that the fluid flow exactly reverses its direction upon the reversal of actuating forces, whereas the tendency of a system to reach an ordered state is by definition irreversible \cite{golestanian_hydrodynamic_2011}. Theoretical models therefore have to take into account higher order effects that break the respective symmetries. These can include a second degree of freedom per cilium \cite{Reichert2005,guirao_spontaneous_2007,niedermayer_synchronization_2008,qian_minimal_2009,uchida_hydrodynamic_2012,man_multisynchrony_2020}, the asymmetric spatial arrangement of cilia \cite{vilfan_hydrodynamic_2006}, a trajectory or driving force with sufficiently broken symmetries \cite{uchida_synchronization_2010,uchida_synchronization_2010-1,saha_pairing_2019, meng_conditions_2021, kanale_spontaneous_2022, uchida_generic_2011, uchida_hydrodynamic_2012,maestro_control_2018}, or a non-linear driving mechanism that, for instance, switches the direction of force when a switch point is reached \cite{wollin_metachronal_2011,elgeti_emergence_2013, guo_bistability_2018, chakrabarti_hydrodynamic_2019, chakrabarti_multiscale_2022}.

\begin{figure}
    \centering
    \includegraphics[width=\linewidth]{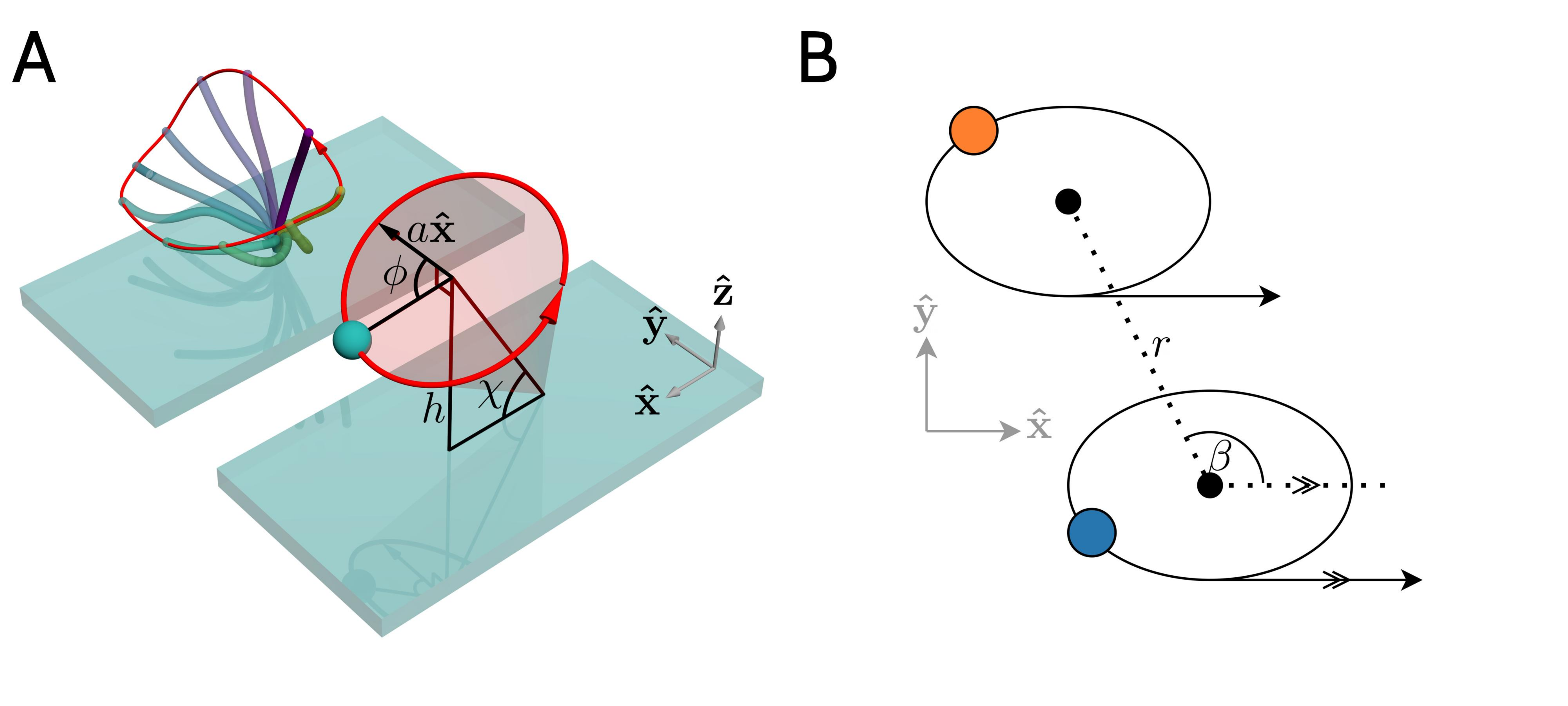}
    \caption{Illustration of the model, showing the parameters used. (A) A realistic cilium motion with the trajectory shown in red. The power stroke (solid blue colour) gives way to a slower recovery stroke along the no-slip surface of the substrate, resulting in net fluid flow in the direction of the power stroke over a cycle. Also shown is one of our model cilia that approximates the realistic motion, with the trajectory shown in red, and relevant quantities indicated. The circular trajectory retains the essential features of a power stroke far from the substrate and a recovery stroke much closer. (B) Definition of $\beta$ and the intercilium distance $r$. The arrows represent the direction of the power stroke of the cilium, occurring at the highest point above the surface. Feathering on lines indicates that they are parallel, so that $\beta$ is the angle between the power stroke and the displacement vector connecting the lattice points of two cilia.}
    \label{fig:figure1}
\end{figure}

When discussing the role of symmetries for ciliary synchronisation, one has to keep in mind that reciprocity manifests itself differently for conservative or dissipative interactions. For conservative forces, Newton's third law states that opposite forces are exerted on both interacting bodies. For hydrodynamic interactions, which are dissipative in their nature, the Lorentz reciprocal theorem \cite{masoud_reciprocal_2019} implies that the force on one body, caused by the motion of a second one with a given velocity, is identical to the force on the second body when the first body moves with the same velocity. Hydrodynamic interactions therefore act on both bodies with the same sign. The interplay of both interaction types is one possibility to facilitate ciliary synchronisation  \cite{niedermayer_synchronization_2008}. In active systems, nonreciprocal interactions can arise where the effect of the interaction on body A differs from that on body B, both in magnitude and direction \cite{soto2014,soto2015,canalejo2019,saha_pairing_2019, saha_scalar_2020, loos_irreversibility_2020, fruchart_non-reciprocal_2021, osat_nonreciprocal_2022}. For example, in the Vicsek model, particles or animals can be affected by other particles in front of them in a different way from those behind them. The orientation of hydrodynamically coupled rotors is a prime example of nonreciprocal coupling that leads to a rich phenomenology including turbulent behaviour via defect proliferation and annihilation \cite{uchida_synchronization_2010}. 

A major open question is related to the scaling with the system size and the role of boundaries of the ciliated region. Recent theoretical work shows that the time needed to reach synchronisation scales quadratically with the number of cilia \cite{solovev_synchronization_2022}. In principle, the metachronal wave vector of the final state is not uniquely determined. However, the basins of attraction of different solutions can greatly differ in size, leading to a strong preference of one state \cite{solovev_synchronization_2022}. Boundaries are often detrimental for synchronisation, because the cilia at the edge have a smaller number of nearest neighbours, which can affect their characteristic frequency as demonstrated in a small 1D row of artificial oscillators \cite{kavre_hydrodynamic_2015}. Boundary effects in a finite system can even lead to the emergence of a chimera state in which the oscillators split up into a coherent and an incoherent population \cite{hamilton_chimera_2017}. The vast majority of theoretical and computational studies focus on systems with periodic boundary conditions as a representation of generic, infinite systems \cite{meng_conditions_2021, elgeti_emergence_2013, uchida_synchronization_2010, uchida_synchronization_2010-1, niedermayer_synchronization_2008, nasouri_hydrodynamic_2016, solovev_synchronization_2022, mannan_minimal_2020, wollin_metachronal_2011}. In nature, periodic circular 1D chains of cilia exist, for instance the oral cilia of \textit{Stentor}~\cite{wan_reorganization_2020} or in starfish larvae~\cite{strathmann_feeding_1971}. However, for topological reasons 2D arrangements of cilia need open boundaries or topological defects, as it is impossible to have a polar field on the topology of a sphere without discontinuities. 

In this paper, we show that the near-field effects between hydrodynamically coupled cilia can lead to an effective nonreciprocal interaction, where cilium A can affect the phase of cilium B more strongly than vice versa. As a result of this nonreciprocity, the metachronal order propagates through the array of cilia with a group velocity, which is not directly related to the velocity of the fluid transport or the phase velocity of metachronal waves. In a finite group of cilia, order then emerges at a boundary and propagates across the group in a time that scales linearly with the system dimension, an order of magnitude faster than an equivalent system without near-field hydrodynamics. We suggest that nonreciprocal coupling is key to understanding the fast emergence of synchronisation in large ciliary carpets. The dynamics of the tissue are then characterised by three independent velocities: the velocity of fluid transport above the surface, the phase velocity of metachronal waves, and the group velocity with which the order propagates. 

\subsection*{Results}

\begin{figure*}
    \centering
    \includegraphics[width=\linewidth]{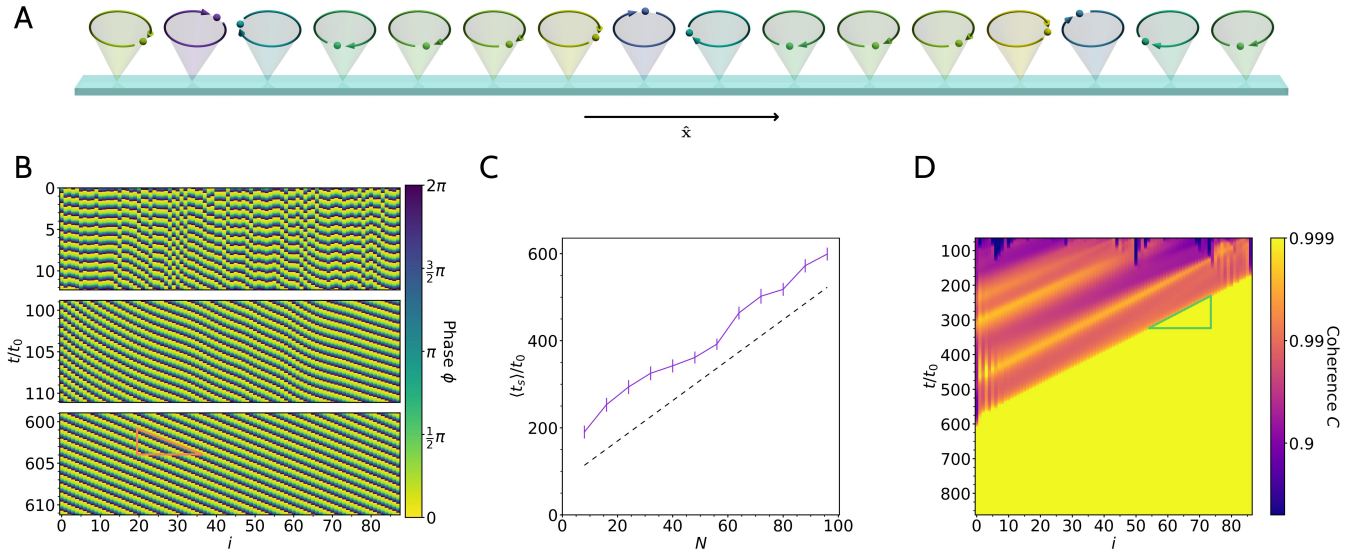}
    \caption{Synchronisation in a one-dimensional row of cilia. 
    (A) A snapshot from the simulation. The colours indicate the phases.
    (B) A kymograph showing the metachronal waves in the system at different times. The simulation starts with random phases, but patches of order quickly assert themselves and give rise to waves that are initially uneven but eventually become completely uniform. The waves travel with the phase velocity $v_\text{ph}$ (orange triangle) in the direction of fluid transport, and are hence symplectic waves. 
    (C) The mean synchronisation time $\left<t_\text{s}\right>$ vs.\ the number of cilia $N$. The mean is calculated by simulating many systems at each size with different random initial phase configurations, and measuring how long it takes to synchronise using a metric based on standard deviation of cilium frequencies. The figure shows that the synchronisation time scales approximately linearly in the system size. 
    Error bars are standard error of the mean, based on $\ge 92$ simulations.
    (D) A kymograph showing the coherence between adjacent pairs of cilia. On this graph each value of $i$ on the abscissa corresponds to the coherence between cilium $i$ and $i+1$.
    Once an ordered patch forms on the right edge it spreads across the row with the group velocity $v_\text{g}$ (green triangle) in the negative x direction. The fact that the synchronisation time is mainly limited by the propagation across the system explains the linear size-dependence in panel (C).}
    \label{fig:figure2}
\end{figure*}

Cilia are long and thin, and beat with a time-irreversible whip-like stroke~\cite{brennen_fluid_1977}. Because of the complexity of the ciliary stroke, its description quickly leads to an intractable number of parameters. We therefore take a simplified approach common to many theoretical models \citep[e.g.,][]{vilfan_hydrodynamic_2006,meng_conditions_2021,kanale_spontaneous_2022} and replace the cilium with a small sphere, pushed along a fixed trajectory by a position-dependent active force. The position of the sphere represents the tip position of a cilium and the active driving force represents the activity of dynein motors of the cilium's axoneme. We thus consider a sphere of radius $b$ moving on a fixed circular trajectory of radius $a$, with its centre a distance $h$ above a surface. The sphere is driven by an internal driving force $F^\text{dr}(\phi)$ and has an internal friction coefficient $\Gamma(\phi)$, both of which act in the tangential direction of the trajectory. The tilt of the trajectory is controlled by an angle $\chi$ such that when $\chi=\pi/2$, the trajectory lies in a plane parallel to the substrate beneath the cilium, shown in Fig.~\ref{fig:figure1}A.

This choice to model the cilia as single spheres on fixed tilted circular trajectories means that we neglect much of the fluid flow driven by the cilium closer to the surface, while preserving the irreversibility of the cilium beat -- essential given the inherent irreversibility of synchronisation. This approximation also replicates the pumping ability of the cilium: when the cilium is closer to the no-slip surface, it produces less fluid flow, and when it is further away it produces more. Over a cycle, the cilium moves a positive net amount of fluid in the direction of its `power stroke'. At distances from the cilium that are several times greater than $h$, this approximation gives almost identical fluid flow to a more detailed treatment of the cilium \cite{vilfan_generic_2012}. In the following, we orient the pumping direction in the positive $x$-direction.

To study synchronisation and the emergence of metachronal waves, we now consider many cilia arranged on a two-dimensional surface. Each point $\mathbf{r}_i = (x_i, y_i, 0)$ represents the position on the substrate directly below the centre of a cilium's trajectory.  A pair of cilia ($i$ and $j$) is characterised by the angle $\beta_{ij}$, which is the angle between the working stroke of cilium $i$ (along the x-axis, Fig.~\ref{fig:figure1})  and the line pointing from $\mathbf{r}_i$ to $\mathbf{r}_j$. These quantities are illustrated in Fig.~\ref{fig:figure1}B.

The position $\mathbf{R}_i$ of the sphere representing cilium $i$ is parameterised as a function of its phase $\phi_i$ following the notation used by Meng et al.~\cite{meng_conditions_2021}:
\begin{equation}
    \mathbf{R}_i(\phi_i) = \mathbf{r}_i + \begin{pmatrix}a\cos \phi_i  \\ a\sin \phi_i \sin \chi  \\ h - a\sin \phi_i \cos \chi  \end{pmatrix}\;.
\end{equation}
To replicate the beating cycle of a cilium, which consists of a fast working stroke followed by a slower sweeping recovery stroke, we introduce a position-dependent force and an internal drag coefficient, which together determine the force-velocity relationship of the active driving force $F^\text{dr}(\phi_i)-\Gamma(\phi_i) v$.  Both can be expanded in a Fourier series as:
\begin{align}
    F^\text{dr}(\phi_i) &= F^\text{dr}_0 \left[ 1 + \sum_n^\infty A_n\cos(n\phi_i) + B_n\sin(n\phi_i) \right], \label{eq:fourier1}\\
    \Gamma(\phi_i) &= \Gamma_0 \left[ 1 + \sum_n^\infty C_n\cos(n\phi_i) + D_n\sin(n\phi_i) \right]. \label{eq:fourier2}
\end{align}
In the following, we only account for terms where $n \le 2$. This simplification is justified, as the first harmonic is known to be essential for synchronisation (and indeed is well-placed to replicate the cilium's beating pattern of a fast power stroke and a slower recovery stroke) but the second harmonic is much more effective at driving the onset of synchronisation and ensuring a more stable synchronised state~\cite{uchida_generic_2011, uchida_hydrodynamic_2012,kanale_spontaneous_2022}.

Due to the linearity of the Stokes flow, the hydrodynamic force  $\mathbf{F}^\text{h}_i$ on a particle is a linear function of the particle's own velocity and the velocities of all other particles it hydrodynamically interacts with. It can be expressed with a generalised friction tensor in the presence of a no-slip boundary, $\mathbf{\Gamma}(\phi_i, \phi_j)$, as $\mathbf{F}^\text{h}_i = - \sum_j \mathbf{\Gamma}(\phi_i, \phi_j) \cdot\mathbf{v}_j$. 
Along with the driving force, which is always tangential to the trajectory, and a perpendicular constraint force $\mathbf{F}^\text{cstr}$ which keeps the particle on the trajectory, the force balance on cilium $i$ states:
\begin{equation}
  \mathbf{F}^\text{dr}(\phi_i) + \mathbf{F}^\text{cstr} - \Gamma(\phi_i)\mathbf{v}_i - \sum_j \mathbf{\Gamma}(\phi_i, \phi_j) \cdot\mathbf{v}_j=0\;.
\end{equation}
By considering only its tangential component (i.e., multiplying the above equation with the tangent vector $\mathbf{t}(\phi_i)$), we obtain the equations of motion for each cilium:
\begin{equation}
  F^\text{dr}(\phi_i) = \Gamma(\phi_t)v_i + \sum_j \mathbf{t}(\phi_i)\cdot \mathbf{\Gamma}(\phi_i, \phi_j)\cdot \mathbf{t}(\phi_j) v_j.\label{eq:eom}
\end{equation}
Here, the velocities are related to the phase derivatives as  $\mathbf{v}_i=(\partial\mathbf{R}_i / \partial\phi_i) {\dot{\phi}}_i$. By solving these equations numerically, we can simulate the evolution of the cilium phases $\phi_i$ over time. In the following, we non-dimensionalise all time units using the time period of an isolated cilium  $t_0$, which can be determined as $t_0=\int_0^{2\pi} \left(\dot \phi_i\right)^{-1} \mathrm{d} \phi_i$ using \eqref{eq:eom} without interacting neighbours.

\begin{figure*}
    \centering
    \includegraphics[width=0.8\textwidth]{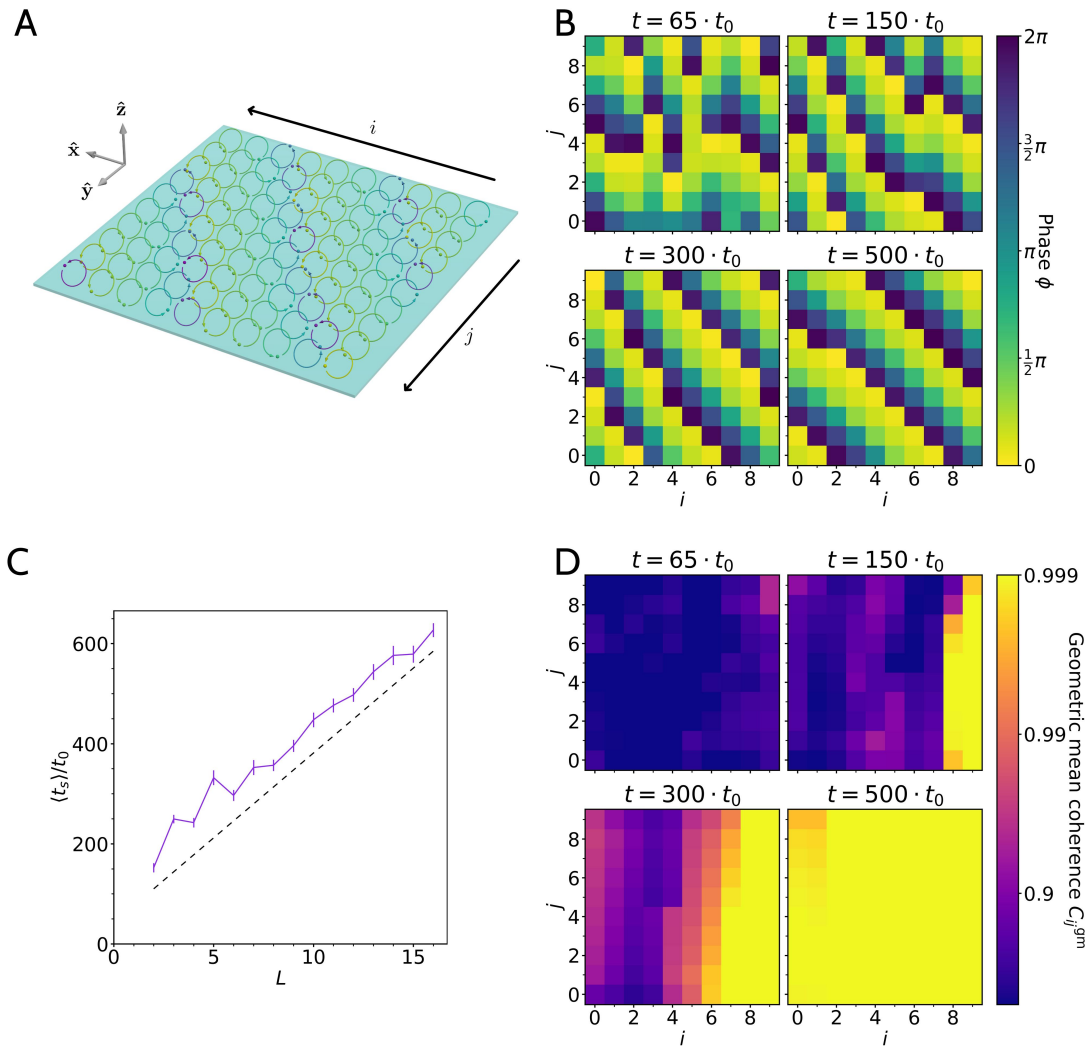}
    \caption{Synchronisation in a two-dimensional square lattice. (A) A schematic of the simulation of the square $L\times L$ lattice at a synchronised state for the specific case $L=10$. The colour of each model cilium indicates its phase, making the order clearly visible. See also Movie S1 for an animated representation. (B) A series of snapshots showing the phases of the cilia in the square lattice, for the specific case of $L=10$ (see movie S2 for a complete time series). (C) The mean synchronisation time $\left<t_\text{s}\right>$ vs. the linear dimension of the system $L$. The mean is calculated by simulating many systems at each size with different random initial phase configurations, and measuring how long it takes for the standard deviation of the cilium frequencies in each system to drop below a certain threshold value. The synchronisation time scales approximately linearly in $L$. 
    Error bars are standard error of the mean, based on $\ge 31$ samples. 
    (D) The geometric mean of the coherence between each cilium and all of its neighbours for a specific simulation with $L=10$. The order emerges on the right side and spreads across the system in negative $x$ direction, leading to the observed linear dependence between the  synchronisation time and the length $L$.}
    \label{fig:figure3}
  \end{figure*}

\subsubsection*{Symmetries}

Before discussing the numerical solutions, it is instructive to consider the symmetries of the system and their effect on synchronisation and formation of metachronal waves. Our model contains the following symmetries:
\begin{enumerate}[label=(\roman*)]
    \item Swapping. Because all cilia are intrinsically equal, the equations of motion stay the same when exchanging two cilia ($\phi_1 \leftrightarrow \phi_2$) and re-arranging them such that $\beta \leftrightarrow \beta +\pi$.
    \item Mirror symmetry. The trajectories of cilia (but not their drag and driving force) are symmetric with respect to $y\leftrightarrow -y$. The equations of motion therefore contain the symmetry $\beta \leftrightarrow \pi - \beta$, $\phi \leftrightarrow \pi - \phi$, $F_0^\text{dr} \leftrightarrow - F_0^\text{dr}$ with the adjustment of the coefficients defined in Eqs.~(\ref{eq:fourier1}, \ref{eq:fourier2}): $(A_n, C_n) \leftrightarrow (-1)^n (A_n, C_n)$ and $(B_n, D_n) \leftrightarrow -(-1)^n (B_n, D_n)$.

    \item Time reversal. Due to the time-reversibility of the Stokes equation, the equations of motion also remain invariant under the transformation $F_0^\text{dr} \leftrightarrow - F_0^\text{dr}$ and $t \to -t$. Because of the time-reversal, a solution that is stable in the original system becomes unstable in the transformed system.

    \item {Without near-field hydrodynamics: axial reflection. If the distance between cilia is sufficient that the near-field hydrodynamic interactions can be neglected ($r \gg h$), the mobility tensor ($\mathbf{M}=\mathbf{\Gamma}^{-1}$) for two particles at a distance $\Delta \mathbf {x}=(\Delta x, \Delta y, 0)$, where $\Delta x = x_j - x_i$ and $\Delta y = y_j - y_i$, can be approximated as \cite{vilfan_hydrodynamic_2006} 
    \begin{equation}
        \mathbf M(\mathbf{x}_i, \mathbf{x}_j) = \frac{3}{2\pi\eta} \cdot \frac{z_i z_j}{\left| \Delta \mathbf{x} \right|^5} \begin{pmatrix}
            (\Delta x)^2 & \Delta x \Delta y & 0 \\
            \Delta y \Delta x & (\Delta y)^2 & 0 \\
            0  & 0  & 0
        \end{pmatrix}.
        \label{eq:eom_ff}
    \end{equation}
    At the same time, the variation of the horizontal positions ($x,y$) of a cilium during a cycle can be neglected such that the motion along the trajectory only affects the vertical distances $z_i$ and $z_j$. The far-field mobility tensor is therefore symmetric with respect to $\beta\leftrightarrow \beta+\pi$.}
\end{enumerate}

The above symmetry properties have bold consequences for the synchronisation. Consider a row of cilia arranged along the $x$ axis, in the direction of pumping. In such a row, the angles $\beta$ can only have values $0$ and $\pi$. 
Without any of the coefficients  that change sign under (ii), i.e., $A_1, C_1, B_2, \ldots$, the motion is symmetric upon the combination of transformations (i), (ii) and (iii). Because the combined transformation contains time reversal which renders a stable solution unstable, no stable states are possible under these assumptions. The notion is consistent with the result in \cite{vilfan_hydrodynamic_2006} if two cilia are arranged along the pumping direction. The existence of a stable solution requires at least one of the terms $A_1, C_1, B_2, D_2, A_3, C_3$, etc. to be nonzero. The same argument also holds for a row of cilia arranged along the $y$ axis (perpendicular to the pumping direction) when the symmetries (ii) and (iii) are employed (see related arguments in \cite{Elfring2009,golestanian_hydrodynamic_2011}).

Without near-field effects (NFEs) in the hydrodynamic coupling, the symmetry property (iv) immediately implies the equivalence of metachronal waves with wave vectors $\mathbf{k}$ and $-\mathbf{k}$, as seen in \cite{meng_conditions_2021}. We therefore expect such systems to show the emergence of multiple long-lived domains with different metachronal wave vectors.

Near-field effects in combination with (for instance) the rotational motion of cilia can break the spatial symmetry, and lead to antisymmetric coupling terms that synchronise the cilia into a state with a non-zero phase difference \cite{vilfan_hydrodynamic_2006,Elfring2009,golestanian_hydrodynamic_2011,solovev_lagrangian_2021,solovev_synchronization_2022-1}. Here, we point out that the interactions are not only asymmetric with respect to the phase difference, but also nonreciprocal with respect to their strength. In a given configuration, the response of cilium $i$ to the phase of cilium $j$ can differ from the response of cilium $j$ to cilium $i$ both in the magnitude and in the phase dependence. This nonreciprocity has profound implications for the emergence of metachronal waves.

\subsubsection*{Synchronisation in one dimension}

We first consider a one-dimensional row of cilia with uniform spacing $d$ and open boundaries such that cilium $i$ is located at position $\mathbf{r}_i=(id,0,0)$ (see Fig.~\ref{fig:figure2}A). This means that $\beta_{ij}=0$ or $\pi$ for every cilium pair $i\ne j$. We used numerical simulations to see how order emerged in the system when the cilia were initialised with random initial phases. 

Figure~\ref{fig:figure1}B shows the phases of the cilia on a kymograph. The initially random phases quickly coalesce into mostly-ordered waves, which slowly become more ordered over time until the waves are completely uniform.  The average time $t_\text{s}$ to reach a synchronised state scales approximately linearly with the system length (Fig.~\ref{fig:figure1}C). We consider the state as synchronised when the standard deviation of all cilium frequencies falls below a fixed threshold. 
The linear dependence can be understood by looking at the signal coherence between adjacent pairs of cilia (Fig.~\ref{fig:figure1}D). The signal coherence is a measure of the degree of linear dependence between two signals, given as a function of the frequency, with values between 0 and 1. For two signals in the time domain $x(t)$ and $x'(t)$, the coherence is calculated as
\begin{equation}
    C_{xx'}(f) = \frac{\left| \Tilde{x}^*(f) \Tilde{x}'(f)  \right|^2}{ \Tilde{x}(f) \Tilde{x}'(f)},\label{eq:coherence}
\end{equation}
where $\Tilde{x}(f)$ and $\Tilde{x}'(f)$ indicate the Fourier transforms of $x(t)$ and $x'(t)$, respectively. For every pair of cilia, we calculate the coherence between $\cos(\phi_i(t))$ and $\cos(\phi_j(t))$ at the frequency with the strongest cross-spectral density between the two signals (i.e., the frequency $f$ that maximises the numerator in \eqref{eq:coherence}).

Random patches of order sometimes emerge and travel against the pumping direction (in this case the pumping direction is rightwards), as the nonreciprocal nature of the hydrodynamic interactions causes the order to expand on one side and be extinguished by the disorder on its other side. However, when an ordered patch occurs close enough to the rightmost edge, there is no disorder to its right to extinguish it, so it spreads throughout the system. This explains why we see that the synchronisation time has a roughly affine relationship with the system length.

\subsubsection*{Synchronisation in two dimensions}

The vast majority of motile cilia are found in two-dimensional bundles on multiciliated cells, where the cells themselves are sparsely distributed~\cite{boselli_fluid_2021}. Hence, we consider a two-dimensional square lattice with side length $L$ and lattice constant $d$ (so that the total number of cilia is $N=L\times L$). We enforce open boundaries, and run a very similar simulation to the one described in the previous section. Figure~\ref{fig:figure3}A shows the lattice, with the cilium trajectories marked according to their phase, rendering the structure of the metachronal wave clearly visible. Figure~\ref{fig:figure3}B shows how the order of the cilia emerges over time: initially there is no correlation between phases, but over time some order emerges, which eventually solidifies into well-ordered metachronal waves.

Figure \ref{fig:figure3}C shows that the synchronisation time scales approximately linearly with the linear dimension of the system $L$ (i.e.\ $\left< t_\text{s}\right> \sim L \sim \sqrt{N}$), just as in the one-dimensional case. This is explained by Fig.~\ref{fig:figure3}D, which illustrates the coherence of each cilium with its neighbours. For each cilium $i$ the value is given by the geometric mean of coherence values with all directly adjacent (not including diagonally adjacent) cilia:
\begin{equation}
    C_{i}^\text{gm} = {\left[ \prod_{j\in \left\{\text{n.n.}\right\}} C\left(\left\{\phi_{i}\right\},\left\{\phi_j\right\}\right) \right]}^{\left(1/N_\text{n.n.}\right)},
\end{equation}
where $C(\left\{\phi\right\}, \left\{\phi'\right\})$ is the coherence, defined over two time series of phases. The resulting graph explains the linearity: the order emerges along one edge and gradually spreads across the entire lattice. Since the limiting factor to synchronisation is the time taken for the order to spread through the length of the system, this time depends on the linear dimension as $L/v_\text{g}$. 

\begin{figure}
    \centering
    \includegraphics[width=\linewidth]{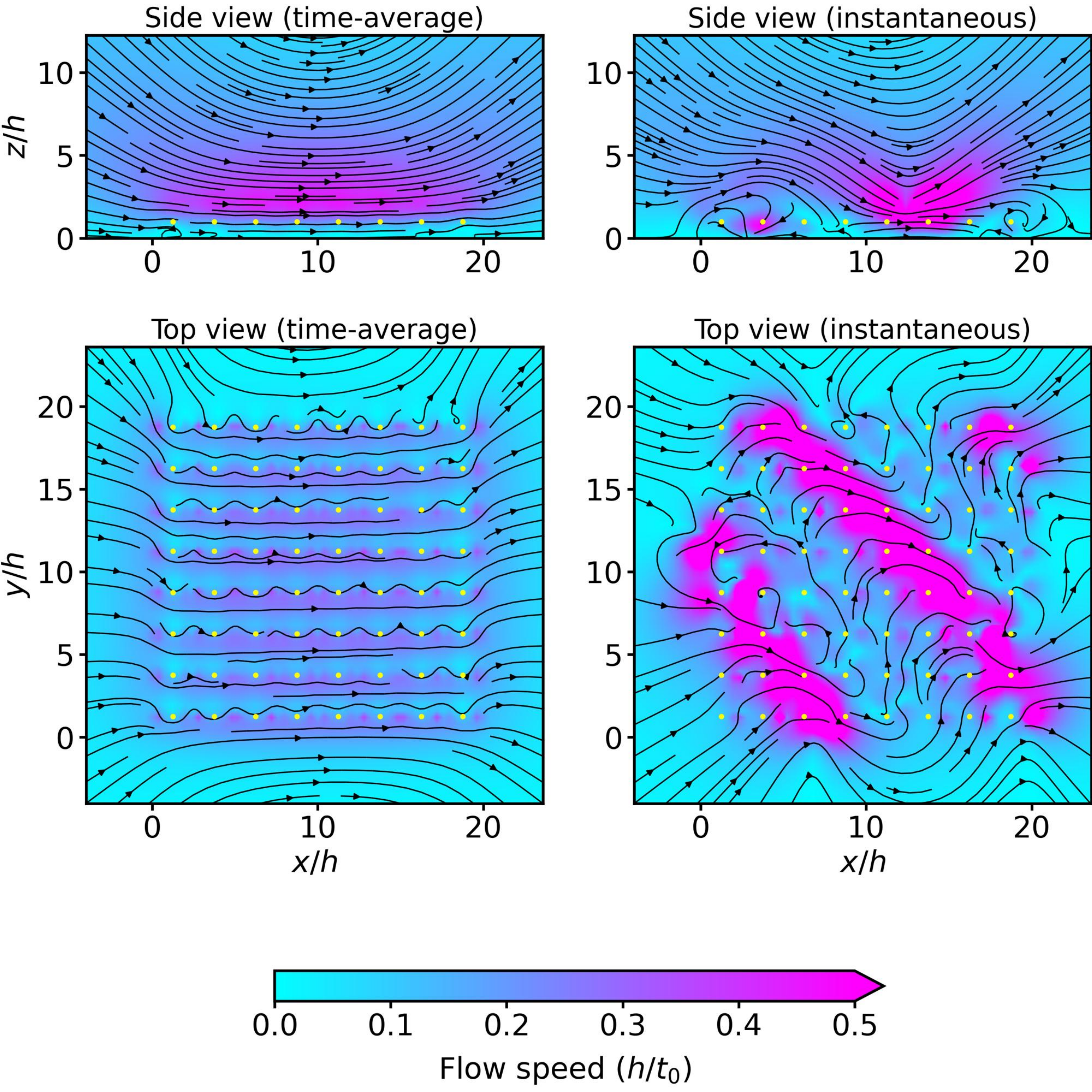}
    \caption{Time-averaged and instantaneous flow in a system of $8\times8$ cilia after reaching a synchronised metachronal state. The background colour indicates the flow speed in units of $h/t_0$ and the yellow dots represent the centre of cilium orbits. The side view corresponds to a vertical cross-section through the middle of the array of cilia ($x/h=10$) and the top view to a slice at $z=h$. The structure of the metachronal wave is clearly visible in the instantaneous flow fields.} 
    \label{fig:figure4}
\end{figure}

The flow field induced by a carpet of cilia that has reached the synchronised state with steady metachronal waves is shown in 
Figure~\ref{fig:figure4}. The time-averaged flows show a region of largely homogeneous flow above the carpet where the fluid is pumped in the positive $x$-direction, the direction of the cilium power stroke. The instantaneous flows, on the other hand, show a periodic structure that follows the movement of metachronal wavefronts.

Although we used a square lattice as an example, the ability of cilia to synchronise is robust against the lattice type and the shape of the arrangement. Similar dynamics is obtained on a hexagonal lattice, as well as on an array with boundaries in the shape of an octagon (Fig.~S1).

\begin{figure*}[t!]
    \centering
    \includegraphics[width=\linewidth]{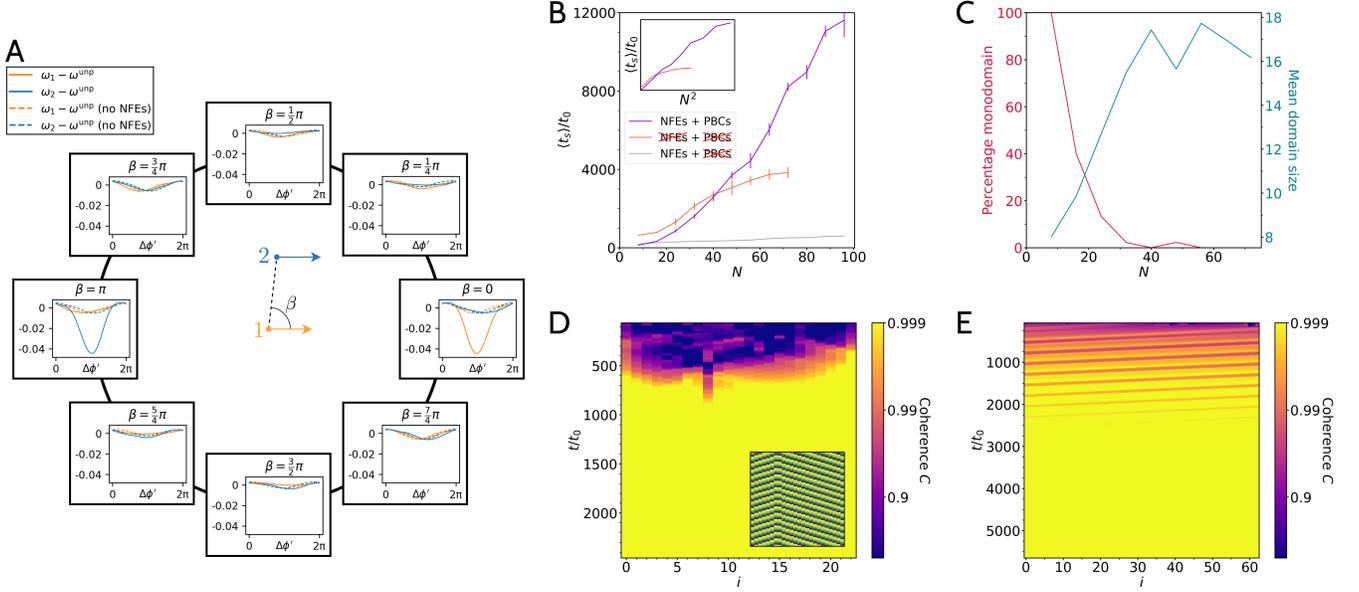}
    \caption{The role of nonreciprocal hydrodynamic interactions and near-field effects in synchronisation. (A)
    The effective angular frequencies $\omega_1$ and $\omega_2$ of two interacting cilia (in dimensionless units) as a function of the adjusted difference $\Delta \phi'$. The cilia are positioned at a fixed distance ($r=2.5h$) in different directions $\beta$. 
    When the cilia are arranged in the $x$-direction ($\beta=0,\pi$) there is a stark difference between $\omega_1$ and $\omega_2$, showing that the interaction is highly nonreciprocal. The nonreciprocity is much weaker when the cilia are arranged $y$ direction ($\beta=\pi/2)$, and the nonreciprocity vanishes entirely when near-field hydrodynamic effects are disabled (dashed lines). (B) The mean time to reach the synchronised state $\left< t_\text{s} \right>$ in a 1D row of $N$ cilia with near-field effects disabled (orange) and with periodic boundary conditions (magenta). The synchronisation time is dramatically longer in both of these cases than in the one-dimensional open boundary case (grey line, data from Fig.~\ref{fig:figure2}C). The inset indicates that the scaling of these synchronisation times is close to $\left< t_\text{s} \right> \sim N^2$. With open boundaries and no near-field effects, however, the synchronisation time reaches a plateau when the final state consists of multiple domains with different wave vectors. 
    Error bars are standard error of the mean, based on 9 samples for the periodic boundary case and $\ge 44$ for the case without near-field effects.
    (C) With near-field effects disabled, the final state typically contains multiple domains with different wave vectors. The red line (left scale) shows the percentage of simulation runs that end in a monodomain state and cyan line (right scale) the average domain size as a function of the system size $N$. (D) Kymograph showing the coherence between adjacent cilia with near-field effects disabled, with the phase kymograph as an inset. Unlike in the case of nonreciprocal coupling (see Fig.~\ref{fig:figure2}D), defects between domains with different wave vectors remain after synchronisation (the example shows one defect). (E) Coherence kymograph of the system with near-field effects and periodic boundary conditions. Defects between coherent regions move with the group velocity, but do not get extinguished at the boundaries, again resulting in a long synchronisation time.}
    \label{fig:figure5}
\end{figure*}

\subsubsection*{Role of nonreciprocity and near-field effects}

Our model shows strong nonreciprocity in the hydrodynamic interactions between cilia. This can be seen by calculating the shift of beating frequencies caused by hydrodynamic interactions, relative to the unperturbed cilium ($\omega-\omega^\text{unp}$). The frequency shifts, averaged over one cycle, are shown in Fig.~\ref{fig:figure5}A as a function of the phase difference $\Delta \phi'$ and the relative position of the two cilia, represented by the angle $\beta$. Nonreciprocity manifests itself as shifts in the beating frequency of the two interacting cilia. The two cilia can experience dramatically different frequency shifts, with very different magnitudes and functional forms. The degree of this nonreciprocity is highly anisotropic, being much greater in the pumping direction than perpendicular to it (Fig.~\ref{fig:figure5}A).

As shown in the section Symmetries, nonreciprocal interactions are not possible when the hydrodynamic interactions are treated in the far-field approximation. In the far-field, the interaction with a cilium at position $\beta$ has to be identical to the interaction with a cilium at the opposite position $\beta + \pi$. 
We demonstrate this by disabling the near-field effects and replacing the off-diagonal elements of the mobility matrix with the approximation given by \eqref{eq:eom_ff}. The resulting frequency shifts (dashed lines in Fig.~\ref{fig:figure5}A) become reciprocal, as they fulfil $\omega_1(\Delta \phi')=\omega_2(-\Delta \phi')$. 

To investigate the role of near-field effects in the formation of metachronal waves, we simulated the dynamics of a row of cilia (analogous to the results in Fig.~\ref{fig:figure2}) with only far-field interactions. The resulting 
synchronisation times are significantly longer (orange line in Fig.~\ref{fig:figure5}B) than with near-field effects (grey line). In small systems, the scaling with size becomes quadratic (inset in Fig.~\ref{fig:figure5}B), whereas we showed them to be linear in the presence of nonreciprocal coupling. However, in larger systems, the synchronisation times saturate, as the final state no longer consists of a uniform metachronal wave. Rather, the system evolves into a long-lived state consisting of multiple domains with distinct wave vectors. An example with two domains, separated by one defect, is shown in Fig.~\ref{fig:figure5}D. The mean domain size and the likelihood that the system evolves into a mono-domain metachronal wave are shown in Fig.~\ref{fig:figure5}C.

To understand the role of open boundaries in our system, we compared the results to the same system with periodic boundary conditions. 
Periodic boundary conditions are typical in other hydrodynamic models of ciliary or flagellar synchronisation~\cite{meng_conditions_2021, elgeti_emergence_2013, uchida_synchronization_2010, uchida_synchronization_2010-1, niedermayer_synchronization_2008, nasouri_hydrodynamic_2016, solovev_synchronization_2022, mannan_minimal_2020, wollin_metachronal_2011, aref_fluid_2021, ringers_novel_2023} when there are many cilia present (though with rare exceptions~\cite[e.g.][]{chakrabarti_multiscale_2022}), as they ensure that no cilia exist at an open boundary which could cause order to break down -- indeed, when such models are subjected to open boundary conditions, they often find only intermittent synchronisation~\cite{wollin_metachronal_2011}.
Our results show that introducing periodic boundaries, while preserving the nonreciprocal coupling, strongly increases the synchronisation timescale, which again scales quadratically with the system size (Fig.~\ref{fig:figure5}B). The reason why periodic boundary conditions become deleterious to synchronisation can be seen in the coherence kymograph in Fig.~\ref{fig:figure5}E. It shows a number of defects, each propagating with the group velocity $v_\text{g}$, that travel periodically across the system, so the system only slowly reaches a coherent state with a single metachronal wave.

\subsection*{Discussion}

In our study we used a strongly simplified model of a cilium. We replaced the cilium with a single particle moving along a tilted circular trajectory. The tilted trajectory breaks the most important symmetry of the cilium, namely that between the power stroke, when the distance to the surface is larger, and the recovery stroke, when the distance is smaller. This asymmetry is at the core of fluid transport, which does not rely on metachronal co-ordination, although the metachronal waves can improve the energetic efficiency of cilia \cite{osterman_finding_2011, elgeti_emergence_2013}. At the same time, the driving force and the internal friction are modulated such that they reproduce a power stroke that is faster than the recovery stroke and also reproduce the fore-aft asymmetry that is present in cilia. The modulation of both parameters represents both the cyclic activity of dynein motors and the variations in the shape of the cilium, which is stretched during the power stroke and bent during the recovery stroke. Unlike theoretical models with fewer broken symmetries \cite{kanale_spontaneous_2022}, our model allows the emergence of metachronal waves that are not directly linked to the fluid transport.

The numerical solution of the model equations takes into account not only the the far-field hydrodynamics, as in \cite{meng_conditions_2021}, but also the near-field effects that become relevant when the size of a cilium becomes comparable to the distance between adjacent cilia. Near-field effects are definitely important in most ciliary systems that show metachronal co-ordination. For example, in \textit{Paramecium} the intercilium distance is in the micrometre range, which is several times less than the cilium length \cite{bouhouche_paramecium_2022}. In respiratory epithelia the distances are even shorter at fractions of a micrometre \cite{sleigh_propulsion_1988}. On the other hand, in \textit{Volvox} colonies, pairs of flagella (one on each cell) are spaced at a distance comparable to their length and still form metachronal waves \cite{brumley_hydrodynamic_2012}. The intermediate densities we chose here allow us to take a generic approach that does not depend on fine details of the trajectory, while qualitatively capturing the near-field interactions. We therefore expect that the magnitude of near-field effects, as well as interactions in general in our study, is underestimated, and that the underlying principles can account for significantly faster synchronisation in natural cilia.

Our main finding is that the near-field effects can make the coupling nonreciprocal. The nonreciprocity goes beyond the asymmetry discussed in \cite{solovev_synchronization_2022}, which implies that two cilia tend to synchronise with a phase difference that depends on their relative orientation. The nonreciprocal magnitude of the interaction means that a cilium tends to follow a neighbour on one side and to entrain the neighbour in the opposite direction. An easy-to-understand mechanism that contributes to nonreciprocity is that the periodically modulated driving force and internal drag make the cilium more susceptible to hydrodynamic interactions in certain parts of the trajectory, which are in turn closer to some neighbours than the others. The nonreciprocal coupling introduces a third direction in the plane, after the direction of fluid transport and the direction of the preferred metachronal wave, which dictates the propagation of order. We therefore refer to it as a group velocity. However, we note that unlike in classical waves in linear media with energy conservation, the group velocity is not related to the phase velocity in a straightforward way (e.g.\ through a dispersion relation). 

Nonreciprocal coupling has two major effects on the formation of metachronal waves. First, it produces robust waves in finite systems with open boundaries. While open boundaries are the standard in real systems, they are detrimental in many models of synchronisation, and also in experimental model systems \cite{kavre_hydrodynamic_2015}. The majority of theoretical works on cilia synchronisation therefore only investigate systems with periodic boundary conditions. In the presence of nonreciprocal coupling, the situation reverses and boundaries help seed the order which then rapidly spreads across the system. With nonreciprocal interactions, it is actually the periodic boundary conditions that significantly slow down the convergence to an ordered metachronal wave. The second major effect of nonreciprocal coupling is that the timescale of metachronal wave formation scales linearly with linear dimension of the system. This holds in both one and two dimensions, due to the linear spreading of order through the system from a boundary. At each system size tested, as long as near-field effects are not suppressed, the system always converges to the same metachronal wavevector regardless of the random initial conditions, meaning that the basin of attraction is effectively as large as the phase space of the system. We have demonstrated that suppressing the near-field hydrodynamic interactions (and therefore the nonreciprocal coupling) gives rise to unfavourable synchronisation time scaling and unpredictable final states with long-living defects remaining.

Our model does not account for non-hydrodynamic interactions which have been shown to be relevant for cilium synchronisation, such as steric effects \cite{chelakkot_synchronized_2021} and basal coupling \cite{liu_transitions_2018}. Because it has been shown that hydrodynamic interactions alone are sufficient to achieve synchronisation \cite{brumley_flagellar_2014}, one can consider these other effects as intercilium coupling to fine-tune the interactions rather than being an absolute requirement. In particular, basal coupling could provide a means to align metachronal waves in order to optimise efficiency \cite{soh_intracellular_2022}. Finally, we neglected any inertial effects which are known to be small compared to viscous forces in systems of cilia. Nevertheless, it is still possible that a small inertial effect can be decisive for synchronisation in situations where other effects cancel out~\cite{theers_synchronization_2013,wei_measurements_2021}.

Our results leave some open questions that could be addressed in future work. For example, in real biological systems there are a great many sources of noise~\cite{gilpin_multiscale_2020}, and at the scale of cilia, noise may be very relevant for synchronisation~\cite{solovev_synchronization_2022-1} so future extensions to our model could examine the role of noise in the motion of the cilia. Additionally, we have assumed that all cilia are of identical lengths, but in reality there can be variation in the lengths of cilia, and some studies have found that this can affect synchronisation \cite{bottier_how_2019}. Similarly, even in healthy humans there are some cilia with structural abnormalities \cite{verra_ciliary_2013}, which means that the influence on synchronisation of nonidentical cilia may be significant. Our circular trajectory retains many key features of the stroke of real cilia, but it is possible that some crucial feature is lost in this simplification, so future work could integrate realistic cilium strokes with elongated cilia. This would also enable a more realistic driving engine for the cilia: in our model the cilia have a time-varying driving force that always points along the tangent of the trajectory, but real cilia are driven by creating shear forces between pairs of dynein tubes that make up the internal structure of the cilium \cite{horani_advances_2018}. It is possible that in the future, artificial or lab-grown cilia may have applications in microfluidic pumping, given the advancing state of the fields of growing artificial lab-on-a-chip cilia \cite{nawroth_stem_2019} and nanoscale artificial cilium production \cite{toonder_microfluidic_2013}, which could offer real-world applications for our work and the future work proposed here.

\section*{Methods}
  \subsubsection*{Fluid flow}

    At the scale of cilia, the behaviour of the fluid flow field $\mathbf{u}$ is well-approximated by the incompressible Stokes equations:
    \begin{align*}
        \eta \nabla^2 \mathbf{u} - \nabla p &= 0, \\
        \nabla \cdot \mathbf{u} &= 0,
    \end{align*}
    where $\eta$ is the fluid dynamic viscosity and $p$ is the pressure.
    
    The hydrodynamic interactions between two particles are calculated using a modified Rotne-Prager approximation with corrections to account for the no-slip fluid boundary on the surface below the cilium. The Rotne-Prager tensor takes into account terms up to the order $\sim r^{-3}$ and is equivalent to averaging the Green's function (Oseen tensor) over the surfaces of both spheres. To take into account the presence of the no-slip boundary at $z=0$, we use the method of images and replace the free-space Green's function by the Blake tensor~\cite{blake_note_1971}, defined as
    \begin{multline}
        \mathbf{M}_{ij}^\text{Blake} = \frac{1}{8\pi\eta} \bigl[ \mathbf{G}^\text{S}(\mathbf{x}_i-\mathbf{x}_j) - \mathbf{G}^\text{S}(\mathbf{x}_i-\mathbf{\Bar{x}}_j) \\
         + 2z_j^2\mathbf{G}^\text{D}(\mathbf{x}_i-\mathbf{\Bar{x}}_j) - 2z_j\mathbf{G}^\text{SD}(\mathbf{x}_i-\mathbf{\Bar{x}}_j) \bigr],
    \end{multline}
    where $\mathbf{x}_k$ is the position of particles $k$, and $\mathbf{\Bar{x}}_k$ is the position of the image of particle $k$ reflected in the no-slip boundary at $z=0$, and where
    \begin{align}
        \mathbf{G}^\text{S}_{\alpha\beta}(\mathbf{r}) &= \frac{\delta_{\alpha\beta}}{r} + \frac{r_\alpha r_\beta}{r^3}, \\
        \mathbf{G}^\text{D}_{\alpha\beta}(\mathbf{r}) &= (1 - 2\delta_{\beta z}) \frac{\partial}{\partial r_\beta} \left(\frac{r_\alpha}{r^3}\right),\\
        \mathbf{G}^\text{SD}_{\alpha\beta}(\mathbf{r}) &= (1 - 2\delta_{\beta z}) \frac{\partial}{\partial r_\beta} G^\text{S}_{\alpha z}(\mathbf{r}).
    \end{align}
    The Rotne-Prager tensor corrected for the no-slip boundary follows by including the leading corrections that result from surface-averaging over each sphere. 
    The non-diagonal terms, describing the interaction between two particles $i\ne j$, can be calculated as
        \begin{equation}
          \mathbf{M}_{ij} = \left( 1 + \frac{a^2}{6} \nabla^2_{\mathbf{x}_i} \right) \left( 1 + \frac{a^2}{6} \nabla^2_{\mathbf{x}_j} \right)  \mathbf{M}_{ij}^\text{Blake}.
        \end{equation}
    Explicit expressions for the elements of the mobility matrix can be found in \cite{vilfan_self-assembled_2010}.

    \subsubsection*{Solving equations of motion} 

    The equations of motion as stated in \eqref{eq:eom} give a complete description of the system. However, they require the knowledge of the many-particle drag matrix $\mathbf{\Gamma}$, whereas the Rotne-Prager approximation gives us the mobility matrix $\mathbf{M}=\mathbf{\Gamma}^{-1}$. Simulating \eqref{eq:eom} directly for $N$ cilia would therefore require the inversion of a $3N\times 3N$ matrix at each simulation step, in addition to solving a linear equation system with $N$ unknowns. 

    To accelerate the numerical solution, we therefore rewrite the equations of motion based on the mobility matrix  $\mathbf{M}(\phi_i,\phi_j)$ which gives the velocity response at the position of cilium $i$ to a force at cilium $j$. We can express the force balance and the hydrodynamic equations with the hydrodynamic force $\mathbf{F}^\text{h}_i$ acting on the cilium:
    \begin{align}
        0 &= \mathbf{t}_i\cdot\mathbf{F}_i^\text{h}(\left\{ j \right\}) + F_i^\text{dr}(\phi_i) - \Gamma(\phi_i)v_i,\label{eq:methodshydrodynamicforces} \\
          \mathbf{v}_i &= - \mathbf{M}(\phi_i,\phi_i)\cdot\mathbf{F}_i^\text{h} - \sum_{j\ne i} \mathbf{M}(\phi_i,\phi_j)\cdot\mathbf{F}_j^\text{h}.
    \end{align}
    By multiplying the first equation with $\mathbf{t}_i$ and inserting it into the second, we can derive a coupled set of $3N$ equations  which allow us to solve for all hydrodynamic force vectors simultaneously (assuming that $\Gamma(\phi)$ is never zero):
    \begin{equation*}
        \left(\mathbf{M}(\phi_i,\phi_i) + \frac{\mathbf{t}_i \mathbf{t}_i^T}{\Gamma(\phi_i)}\right)\cdot\mathbf{F}_i^\text{h} + \sum_{j\ne i} \mathbf{M}(\phi_i, \phi_j)\cdot\mathbf{F}_j^\text{h} = -\frac{F^\text{dr}(\phi_i)}{\Gamma(\phi_i)} \mathbf{t}_i.
    \end{equation*}
    In the above equation system the first term describing the self-interaction of cilium $i$ is always dominant, whereas the second term describing the hydrodynamic interactions between cilia is weaker and can be treated in a perturbative way. In matrix form the equation is always block-diagonally dominant, which means that it can be solved efficiently using an adapted Successive Over-Relaxation (SOR) algorithm~\cite{young_iterative_1971} that works on 3$\times$3 blocks rather than individual elements. In the initial time step of the simulation, we use the solution to the purely diagonal matrix equation as the first iteration, but in subsequent step it is more efficient to start iterating with the solution of the previous step. In this way only a very small number of iterations ($N_\text{it}=3$) is required to converge to remarkably good accuracy with a relative error $\varepsilon <10^{-6}$.
    Once the hydrodynamic forces have been obtained, they can be substituted back into \eqref{eq:methodshydrodynamicforces} to find the cilium speeds $v_i$, and this can be trivially transformed into the time derivatives of the phases ${\dot\phi}_i$. 
        
    \subsubsection*{Numerical integration}

    The phase of each cilium is updated using the standard Runge-Kutta method (RK4). Unlike implicit methods, Runge-Kutta algorithms require a single calculation of the hydrodynamic forces at each timestep, which is by far the most computationally demanding simulation step. The timestep used was approximately $0.001 \, t_0$.

    \subsubsection*{Quantifying synchronisation}

    To determine whether the entire system has reached a synchronised state, we find the average frequencies of each cilium in a moving window of 50 time periods. We take the standard deviation of these frequencies to be the order parameter of the system.

    When considering pairs of cilia, as in Figs.~\ref{fig:figure2}C,~\ref{fig:figure3}D, and~\ref{fig:figure5}D-E, standard deviations were less useful. Instead, the signal coherence was computed using the phases of adjacent pairs of cilia using Welch's method~\cite{welch_use_1967}, with a moving window in the time domain representing approximately 50 unperturbed cilium cycles. In the 2D case, we instead used the geometric mean of the coherence with all neighbouring cilia.

    \subsubsection*{Uniform phase angle}

    In Fig.~\ref{fig:figure5} we used a transformed phase difference $\Delta\phi' = \phi_2' - \phi_1'$. These angles have the property that for a single isolated cilium, $\dot\phi'$ is constant. The transformed phase can be derived from the original phase angle using
    \begin{equation}
        \phi'(\phi) = \frac{2\pi}{t_0} \int_0^{\phi} \frac{1}{\dot\phi(\phi'')} \, \mathrm{d}\phi'',
    \end{equation}
    where all quantities on the right hand side are for an isolated cilium.

    \subsubsection*{Periodic boundary conditions}

    When considering the effect of cilium $j$ on cilium $i$, only the closest instance of $j$ was considered. In simple terms, if $j$ were right next to $i$, then we would proceed in the same way as if we had no periodic boundaries. However, if $j$ were more than half of the system length away from $i$, then we would instead consider a copy of $j$ translated by the system length, putting it closer to $i$. Since the mobility tensor decays quickly along the surface as $1/r^3$, neglecting the distant cilia does not have any effect on the results.

    \subsubsection*{Numerical parameters}

    In all simulations, we took the lattice constant to be $d=2.5h=2.5a$, and $b=a/10$. $\chi$ was always $\pi/6$ and we used $A_1=-0.55$, $A_2=-0.2$, $B_1=-0.2$, $B_2=0.35$, $C_1=0.3$, $C_2=-0.4$, $D_1=-0.1$, and $D_2=-0.55$. These parameters give a fast working stroke and a slower recovery stroke which break the fore-aft symmetry, consistent with the behaviour of real cilia. The slowest part of the stroke is just before the lowest part of the recovery stroke, where the cilium would be curling up and the tip would therefore be travelling at its minimum speed.

    \begin{acknowledgements}

      This work has received support from the Max Planck School Matter to Life and the MaxSynBio Consortium, which are jointly funded by the Federal Ministry of Education and Research (BMBF) of Germany, and the Max Planck Society. A.V. acknowledges support from the Slovenian Research Agency (grant no. P1-0099)
    \end{acknowledgements}
    \bibliography{ciliaabbrev}

\clearpage
\widetext

\setcounter{figure}{0}
\renewcommand\thefigure{S\arabic{figure}}
    \begin{figure}
\centering
\includegraphics[width=0.6\textwidth]{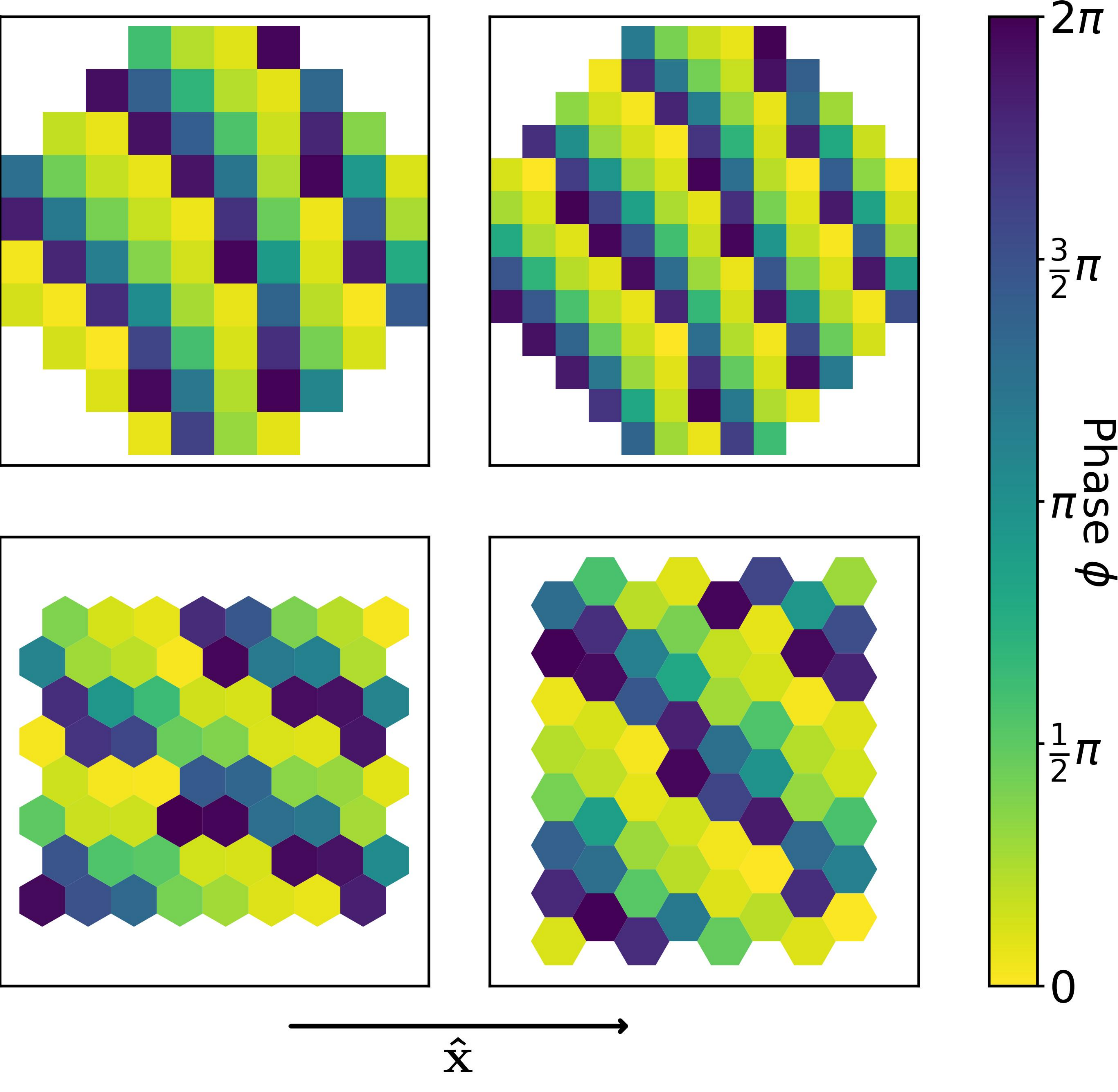}
\caption{Emergent metachronal waves on different lattices, with pumping direction in the x-direction (indicated). (A) 64 cilia arranged on a square lattice forming an octagon. (B) 144 cilia in the same arrangement. (C) 64 cilia on a hexagonal lattice, oriented such that one base vector is aligned with the direction of the power stroke. 
  (D) as in (C), but with a lattice that is rotated by $90^\circ$. }
\label{figS1}
\end{figure}

\begin{figure}
\centering
\includegraphics[width=.5\textwidth]{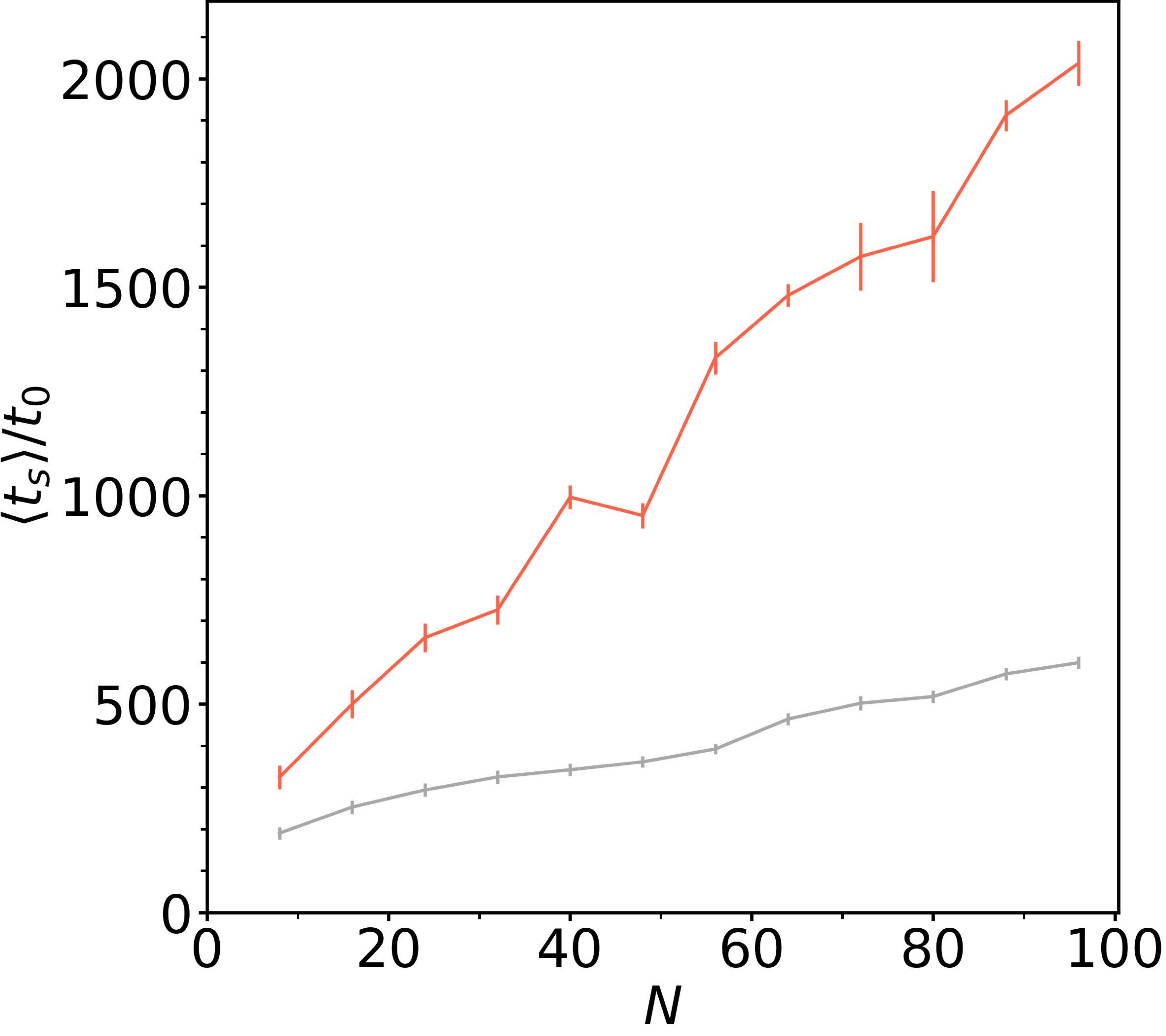}
\caption{Synchronisation times on a 1D lattice. The orange line shows cilia arranged along the y-axis (i.e. in the direction of weaker nonreciprocity). The grey line shows cilia arranged along the x-axis, as in Fig.~2C (main text). Synchronisation times still scale linearly when cilia are arranged in the y-direction, but are higher by some numerical factor, as expected from the weaker (but still present) nonreciprocity.}
\end{figure}

\end{document}